\documentclass{PoS}

\usepackage{float} 
\usepackage[caption=false]{subfig} 

\let\OLDthebibliography\thebibliography
\renewcommand\thebibliography[1]{
  \OLDthebibliography{#1}
  \setlength{\parskip}{-0pt}
  \setlength{\itemsep}{-3pt}
\footnotesize
}

\title{Spin Physics with a fixed-target experiment at the LHC}

\ShortTitle{Spin Physics with a fixed-target experiment at the LHC}

\author{\speaker{M.G. Echevarria}$^{1}$ 
S.J. Brodsky$^{2}$, 
G. Cavoto$^{3}$, 
C. Da Silva$^{4}$,
F. Donato$^{5}$,
E.G. Ferreiro$^{6}$, 
C. Hadjidakis$^{7}$,
I. H\v rivn\'ac\v ov\'a$^{7}$,
D. Kiko\l a$^{8}$, 
A. Klein$^{4}$,
A. Kurepin$^{9}$, 
A. Kusina$^{10}$,
J.P. Lansberg$^{7}$, 
C. Lorcé$^{11}$,
F. Lyonnet$^{12}$,
Y. Makdisi$^{13}$,
L. Massacrier$^{7}$, 
S. Porteboeuf$^{14}$,
C. Quintans$^{15}$,
A. Rakotozafindrabe$^{16}$, 
P. Robbe$^{17}$,
W. Scandale$^{18}$, 
I. Schienbein$^{19}$,
J. Seixas$^{15,20}$, 
H.S. Shao$^{21}$,
A. Signori$^{22}$,
N. Topilskaya$^{9}$,
B. Trzeciak$^{23}$,
A. Uras$^{24}$,
J. Wagner$^{25}$, 
N. Yamanaka$^{7}$,
Z. Yang$^{26}$, 
A. Zelenski$^{27}$
\\
\tiny{
\llap{$^{1}$} INFN, Sezione di Pavia, Via Bassi 6, 27100 Pavia, Italy\\
E-mail: \email{mgechevarria@pv.infn.it}\\
\llap{$^{2}$} SLAC National Accelerator Laboratory, Stanford University, Menlo Park, USA\\
\llap{$^{3}$} ``Sapienza'' Università di Roma, Dipartimento di Fisica \& INFN, Sez. di Roma, P.le A. Moro 2, 00185 Roma, Italy\\ 
\llap{$^{4}$} LANL, P-25, Los Alamos National Laboratory, Los Alamos, NM 87545, USA\\
\llap{$^{5}$} Turin University, Department of Physics, and INFN, Sezione of Turin, Turin, Italy\\
\llap{$^{6}$} Dept. de Física de Partículas, USC, Santiago de Compostella, Spain\\
\llap{$^{7}$} IPNO, CNRS/IN2P3, Univ. Paris-Sud, Université Paris-Saclay, Orsay, France\\
\llap{$^{8}$} Faculty of Physics, Warsaw University of Technology, Warsaw, Poland\\
\llap{$^{9}$} Institute for Nuclear Research, Russian Academy of Sciences, Moscow, Russia\\
\llap{$^{10}$} Institute of Nuclear Physics Polish Academy of Sciences, PL-31342 Krakow, Poland\\
\llap{$^{11}$} CPhT, Ecole Polytechnique, CNRS, Université Paris-Saclay, Palaiseau, France\\
\llap{$^{12}$} Southern Methodist University, Dallas, TX 75275, USA\\
\llap{$^{13}$} Brookhaven National Laboratory, Collider Accelerator Department\\
\llap{$^{14}$} Université Clermont Auvergne, CNRS/IN2P3, LPC, F-63000 Clermont-Ferrand, France \\
\llap{$^{15}$} LIP, Av. Prof. Gama Pinto, 2, 1649-003 Lisboa, Portugal \\
\llap{$^{16}$} IRFU/DPhN, CEA Saclay, 91191 Gif-sur-Yvette Cedex, France\\
\llap{$^{17}$} LAL, Univ. Paris-Sud, CNRS/IN2P3, Université Paris-Saclay, Orsay, France\\
\llap{$^{18}$} CERN, European Organization for Nuclear Research, 1211 Geneva 23, Switzerland\\
\llap{$^{19}$} LPSC, Université Grenoble-Alpes, CNRS/IN2P3, 38026 Grenoble, France\\
\llap{$^{20}$} Dep. Fisica, Instituto Superior Tecnico, Av. Rovisco Pais 1, 1049-001 Lisboa, Portugal\\
\llap{$^{21}$} Laboratoire de Physique Théorique et Hautes Énergies (LPTHE), UMR 7589, Sorbonne Université et CNRS, 4 place Jussieu, 75252 Paris Cedex 05, France \\
\llap{$^{22}$} Physics Division, Argonne National Laboratory, 9700 S. Cass Avenue, Lemont, IL 60439 USA\\
\llap{$^{23}$} Institute for Subatomic Physics, Utrecht University, Utrecht, The Netherlands\\
\llap{$^{24}$} IPNL, Université Claude Bernard Lyon-I, CNRS/IN2P3, Villeurbanne, France\\
\llap{$^{25}$} National Centre for Nuclear Research (NCBJ), Hoza 69, 00-681, Warsaw, Poland\\
\llap{$^{26}$} CHEP, Department of Engineering Physics, Tsinghua University, Beijing, China\\
\llap{$^{27}$} Brookhaven National Laboratory, Collider Accelerator Department}
}

\abstract{

The multi-TeV proton and ion beams of the LHC would allow for the most energetic fixed-target experiment ever.
In particular, $pp$, $p$d and $p$A collisions could be performed at $\sqrt{s_{NN}}$ = 115~GeV, as well as Pb$p$ and PbA collisions at $\sqrt{s_{NN}}$ = 72~GeV, in a parasitic way by making use of the already existing LHCb and ALICE detectors in fixed-target mode.
This would offer the possibility to carry out a ground-breaking physics program, to study the nucleon and nuclear structure at high $x$, the spin content of the nucleon and the phases of the nuclear matter from a new rapidity viewpoint.
In this talk I focus on the spin physics axis of the full program developed so far by the AFTER@LHC study group.
}

\FullConference{23rd International Spin Physics Symposium - SPIN2018 -\\
		10-14 September, 2018\\
		Ferrara, Italy}

\begin{document}

\section{Introduction}

The multi-TeV LHC beams could offer several unique advantages in fixed-target mode~\cite{Brodsky:2012vg,Lansberg:2012kf,Rakotozafindrabe:2012ei,Rakotozafindrabe:2013au,Kurepin:2015jka,Massacrier:2015qba,Kikola:2017hnp,Trzeciak:2017csa,Hadjidakis:2018ifr}:
access to the far backward rapidity region, which remains to be explored with hard reactions;
polarization of the target, which allows for the measurement of single-spin asymmetries;
possibility of using several target types, like deuteron and $^3$He;
an energy range between top SPS and RHIC energies for lead-induced collisions; and
an outstanding luminosity thanks to the high density of the targets and the large LHC beam fluxes.

Given these advantages of the fixed-target mode compared to the collider mode, we have developed a full physics program to be carried out at the LHC in fixed-target mode, based on three main research axes:
i) the high-momentum-fraction ($x$) frontier in nucleons and nuclei, with a specific emphasis on the gluon and heavy-quark distributions, the transition between the inclusive and exclusive regimes of QCD and the implications for astroparticle physics including Ultra-High Energy cosmic neutrinos;
ii) the spin content of the nucleons, with a focus on single transverse spin asymmetries (STSAs) and azimuthal asymmetries from correlations between the spin and the transverse momenta of the partons/hadrons;
iii) the ultra-relativistic heavy-ion collisions in a new rapidity and energy domain, between SPS and RHIC energies, with heavy-flavour observables (including quarkonia) as well as identified light hadrons through a rapidity scan down to the target rapidity.

\begin{figure}[t!]
\centering
%
\includegraphics[width=0.8\textwidth]{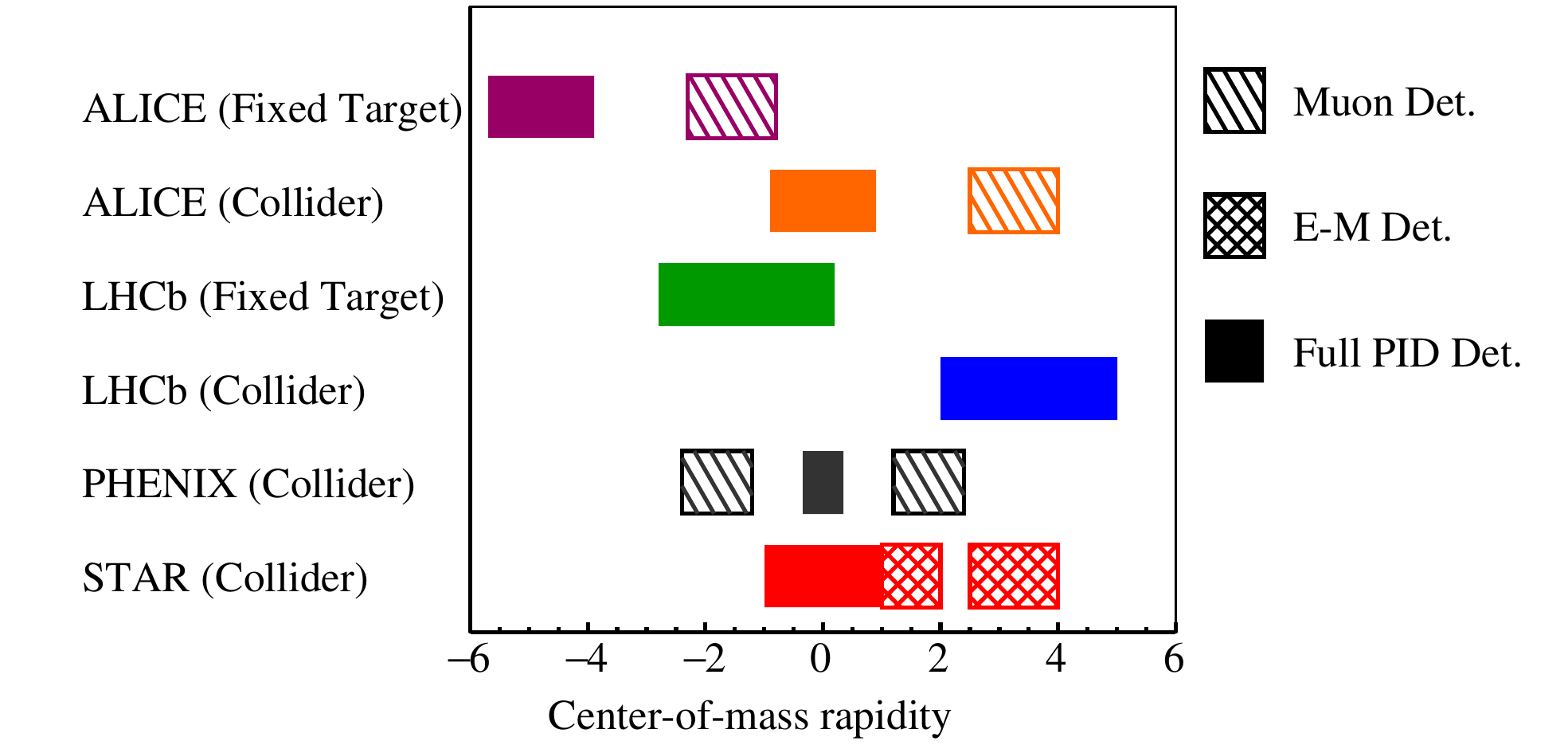}    
%
\caption{\emph{Comparison of the kinematical coverages of the ALICE and LHCb detectors in fixed target and collider modes, with STAR and PHENIX. 
``Full PID Det.'' labels detectors with particle identification capabilities, ``E-M Det.'' an electromagnetic calorimeter and ``Muon Det.'' a muon detector. 
Target position is considered at the nominal Interaction Point.}}
\label{fig:rap}
\end{figure}

7~TeV protons on a fixed target release a center-of-mass-system (cms) energy $\sqrt{s_{NN}}=114.6$~GeV, while for 2.76 TeV Pb beam $\sqrt{s_{NN}}=72$~GeV.
This results in a rapidity shift as large as 4.8 (4.3 for Pb beam).
This large rapidity boost implies that the backward rapidity region is accessible by using the already existing ALICE or LHCb detectors, whose nominal acceptances are then shifted to negative $y_{cms}$ (see fig.~\ref{fig:rap}). 

Different technological implementations are currently under investigation to perform fixed-target experiments at the LHC:
by letting the full LHC beam go through a (possibly polarised) gas target in the LHC beam pipe;
by extracting halo particles by means of a bent-crystal deflector onto a target positioned inside the beam pipe or outside the beam pipe with a dedicated beam line;
or by placing a wire/foil target intercepting the faint beam halo in the beam pipe.
Technical details can be found e.g. in the recent full report of the AFTER@LHC study group~\cite{Hadjidakis:2018ifr}.

\section{3D structure of the nucleon}

The spin part of the fixed-target physics program at the LHC bears on two main pillars: STSA measurements with a polarized target and azimuthal-modulation studies. 
Both allow one to investigate the tri-dimensional (3D) and spin structure of partons confined in hadrons, and help unravel how quarks and gluons are bound in a spin-$1/2$ nucleon.
In particular, STSA measurements in Drell-Yan (DY) process will contribute to the worldwide effort towards the verification of the Sivers-asymmetry~\cite{Sivers:1989cc} sign change~\cite{Collins:2002kn,Brodsky:2002cx} between DY and semi-inclusive deep-inelastic scattering (SIDIS).
The corresponding extractions in the gluon sector, given their expected precision, would simply be ground-breaking. 
In addition, looking for azimuthal modulations in DY and pair-particle production will provide novel ways to study quark-hadron momentum-spin correlations.
Here we review a selection of the projections performed by the AFTER@LHC study group~\cite{Hadjidakis:2018ifr}.

\begin{figure}[t!]
\centering
%
\includegraphics[width=0.48\textwidth]{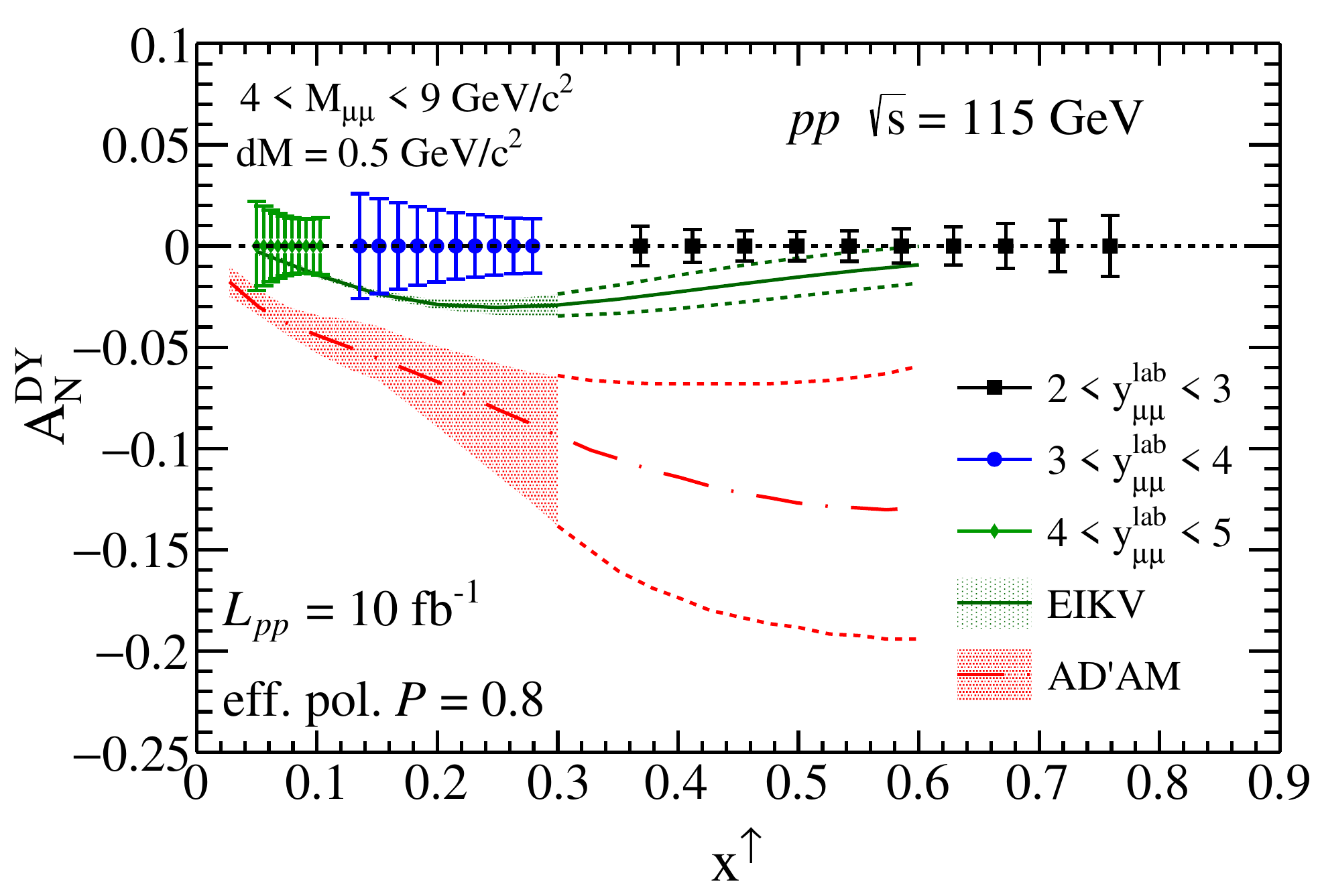} 
\includegraphics[width=0.48\textwidth]{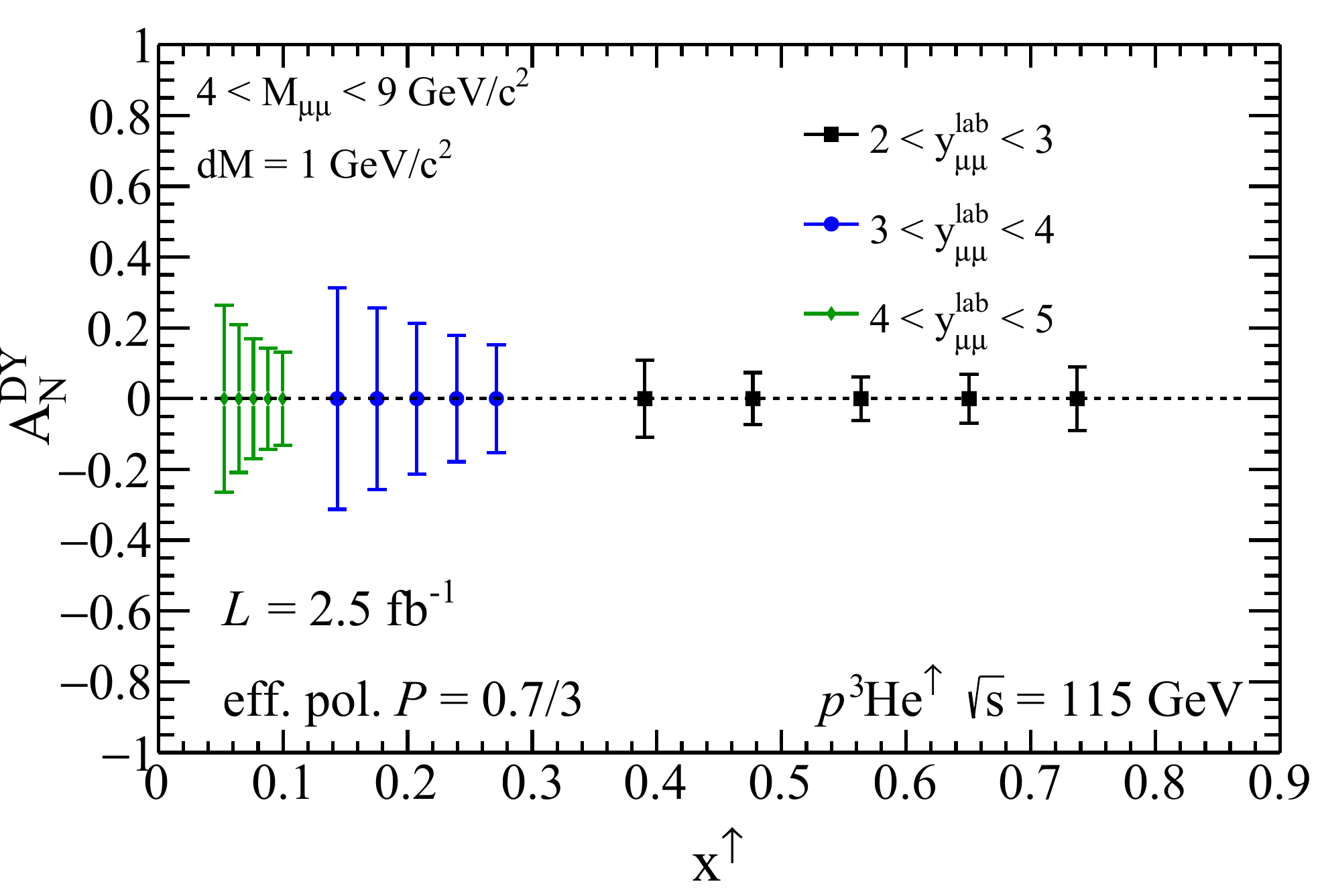}
%
\caption{\emph{(left) Two predictions (AD'AM~\cite{Anselmino:2015eoa} and EIKV~\cite{Echevarria:2014xaa}) of the DY $A_N$ at LHCb in the fixed-target mode, compared to the projected precision of the measurement~\cite{Kikola:2017hnp}. 
Bands are filled in the region where the fits use existing SIDIS data and hollow where they are extrapolations.
(right) Similar projections for the DY $A_N$ as a function of $x^{\uparrow}$ in $p+^{3}$He$^{\uparrow}$ collisions~\cite{Kikola:2017hnp}.
}}
\label{fig:spin1}
\end{figure}

Fig.~\ref{fig:spin1}~(left) shows two theoretical predictions of the DY STSA ($A_N$) compared to the projected statistical precision using the LHCb detector in the fixed-target mode.
Clearly this measurement can put strict constraints on the quark Sivers effect and the related 3-parton correlation functions~\cite{Efremov:1981sh,Qiu:1991pp}, help discriminate among different approaches, and test the time-reversal symmetry of QCD, realized through the relevant initial/final state interactions for DY and SIDIS.
It is clearly crucial to be able to perform measurements in the rapidity range $2 < y_{\rm lab} <3$ in order to access the still unmeasured high $x^\uparrow$ region~\cite{Kikola:2017hnp}.
In Fig.~\ref{fig:spin1}~(right) we show a similar projection but in $p+^{3}$He$^{\uparrow}$ collisions.
This measurement could not only complement the ones carried out at JLab in the last two decades, but improve them by offering more precision and the possibility to constrain the Sivers function in a neutron, shedding some light on its isospin dependence.

\begin{figure}[t!]
\centering
%
\includegraphics[width=0.32\textwidth]{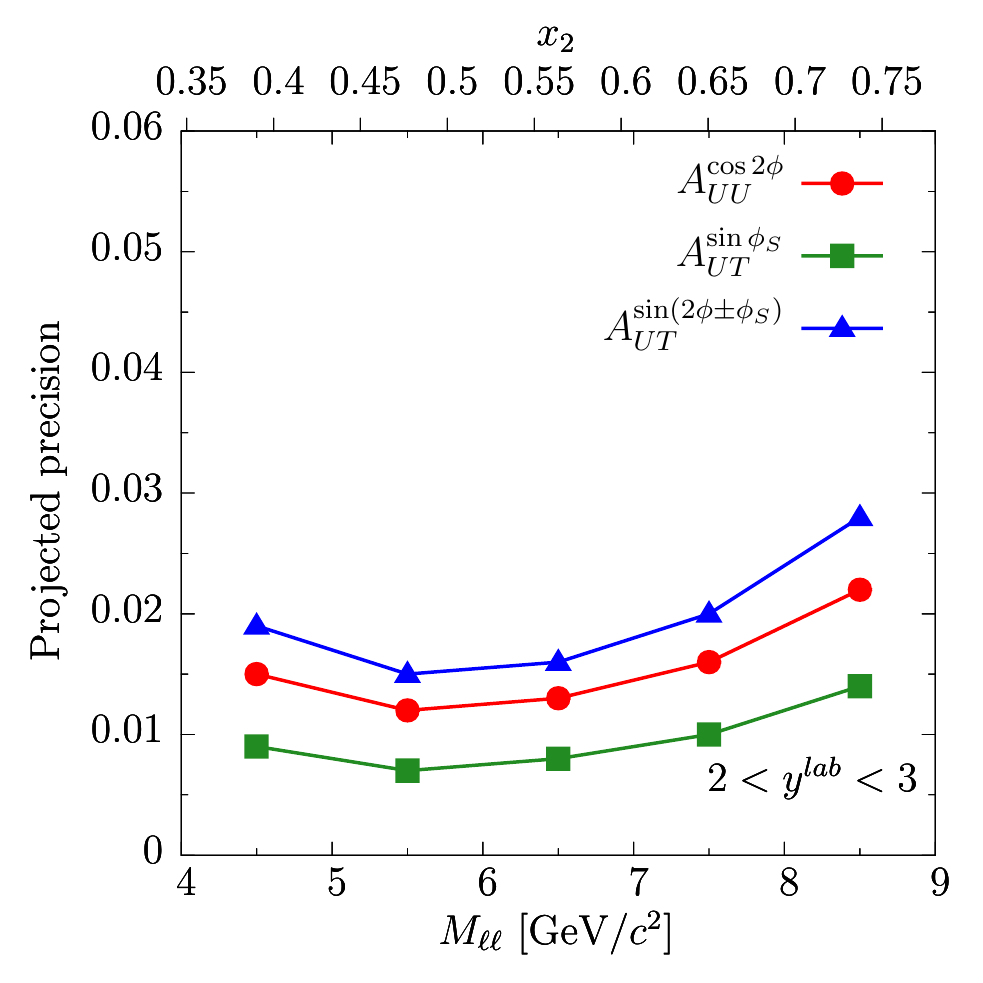}    
\includegraphics[width=0.32\textwidth]{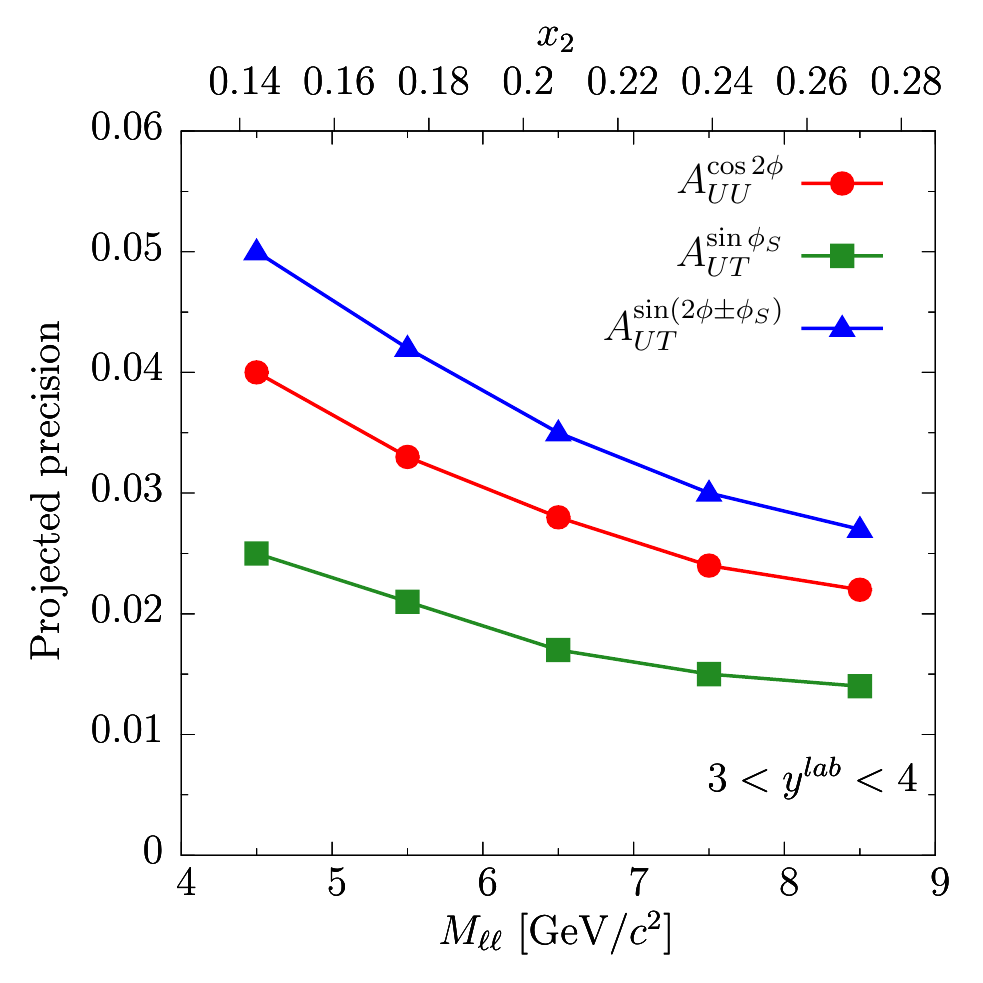}    
\includegraphics[width=0.32\textwidth]{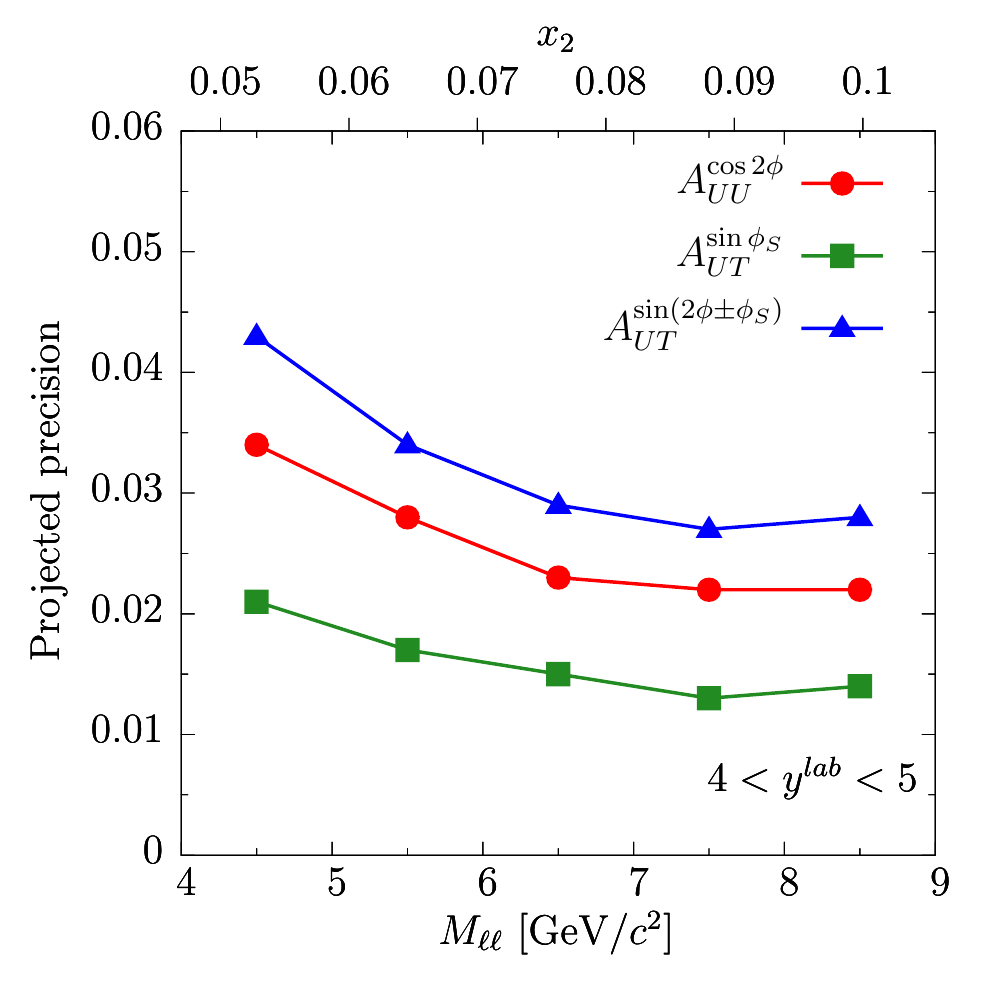}    
%
\caption{\emph{Projected statistical uncertainty on asymmetries in DY production with LHCb (the rapidity is integrated over, as well as the mass in bins of $dM=1$~GeV/$c^2$).}}
\label{fig:spin2}
\end{figure}

In Fig.~\ref{fig:spin2} we show the projected statistical precision for several other asymmetries in DY production, which is expected to be as good as a few percent.
They would for the first time offer the opportunity to constrain even less known transverse-momentum distributions, such as $h_1^q$, $h_1^{\perp q}$ and $h_{1T}^{\perp q}$, which encode other quark-hadron momentum-spin correlations, and which are indirectly related to the quark orbital angular momentum.

\begin{figure}[t!]
\centering
%
\includegraphics[width=0.48\textwidth]{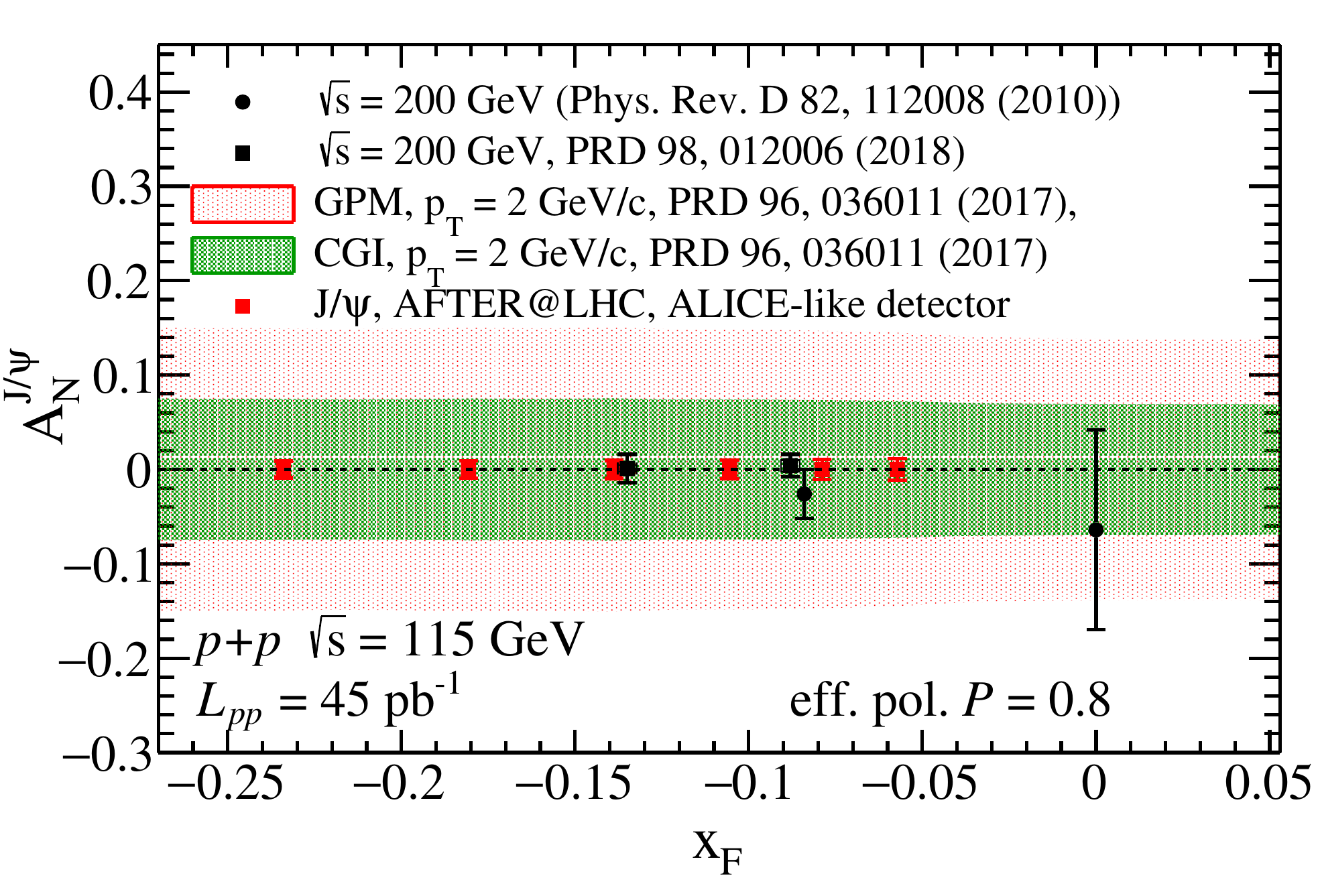}    
%
\caption{\emph{$J/\psi$ $A_N$ projected precision with the ALICE detector compared to existing measurements~\cite{Adare:2010bd,Aidala:2018gmp}.}}
\label{fig:spin3}
\end{figure}

A major strength of the LHC fixed-target mode is the large production rates for open heavy-flavour mesons and quarkonia (roughly $10^6$ $\Upsilon$ and $10^9$ $J/\psi$ in $p$H collisions for a single year of data taking~\cite{Massacrier:2015qba,Feng:2015rka}).
These processes are very useful probes to precisely access and constrain the gluon Sivers effect, still essentially unknown~\cite{Boer:2015vso}.
For instance, Fig.~\ref{fig:spin3} shows the projected statistical precision with ALICE in fixed-target mode for $J/\psi$ $A_N$, compared to existing measurements~\cite{Adare:2010bd,Aidala:2018gmp} and theoretical predictions~\cite{DAlesio:2017rzj}.

Photoproduction in hadron-hadron interactions~\cite{Baltz:2007kq} can also be accessed with ultraperipheral collisions in the fixed-target mode~\cite{Lansberg:2015kha,Lansberg:2018fsy}. 
These exclusive reactions can constrain generalised parton distributions~\cite{Diehl:2003ny, Belitsky:2005qn,Radyushkin:1997ki,Ji:1996nm,Mueller:1998fv}, which parametrize the three-dimensional structure of hadrons in position space, and which are related to parton orbital angular momentum.

Finally, at the LHC in fixed-target mode one can also access the distributions of longitudinally polarised (anti)quarks inside hadrons, which are not well known~\cite{Nocera:2014gqa}, through the measurement of the longitudinal spin transfer $D_{LL}$ from a longitudinal polarised target to $\Lambda$ or $\bar{\Lambda}$ hyperons.
The understanding of the possible asymmetry in the distributions of polarised strange and anti-strange quarks is still an open and very intriguing question in hadronic physics.
In addition, similar measurements can be performed for the transverse spin transfer $D_{TT}$ to hyperons, giving access to the integrated quark transversity.
This quantity, also called the nucleon tensor charge, is a very useful input for searches of new physics beyond the Standard Model.

\section{Summary}

A fixed-target experiment at the LHC would greatly extend its physics capabilities, offering many opportunities to study the nucleon/nuclear structure at high $x$, the properties of nuclear matter under extreme conditions in heavy-ion collisions, and the nucleon 3D/spin decomposition in terms of partonic degrees of freedom, which has been the topic of this talk.
Extensive theoretical works have contributed to the development of a full physics program for a fixed-target experiment at the LHC, using both the multi-TeV proton and ion beams~\cite{Ceccopieri:2015rha,Feng:2015rka,Kanazawa:2015fia,Anselmino:2015eoa,Brodsky:2015fna,Boer:2015vso,Liu:2012vn,
Boer:2012bt,Lansberg:2015lva,Lansberg:2015kha,Arleo:2015lja,Vogt:2015dva,Begun:2018efg,Karpenko:2018xam,
Brodsky:2018zdh,Ceccopieri:2015rha,Goncalves:2018htp,Goncalves:2015hra}.
Several projection studies, based on the performances of ALICE and LHCb detectors in fixed-target mode~\cite{Hadjidakis:2018ifr,Kikola:2017hnp,Trzeciak:2017csa,Massacrier:2015qba,Kikola:2015lka,Kurepin:2015jka,Ulrich:2015uwb}, clearly show that unprecedented precise measurements are at reach, both on quark and gluon sensitive probes such as Drell-Yan, open heavy-flavour and quarkonium production.

{\bf \emph{Acknowledgements.}}
MGE is supported by the Marie Sk\l odowska-Curie grant GlueCore (grant agreement No. 793896).

\bibliographystyle{Science}
\bibliography{references}

\end{document}